\documentclass[aps,twocolumn,showpacs,floatfix,superscriptaddress]{revtex4}
\usepackage{amsmath}
\usepackage{amsfonts}
\usepackage{amssymb}
\usepackage{epsfig}
\usepackage{graphicx}

\begin{document}

\title{Squeezed states of quadratically kicked nanomechanical oscillators}

\author{G. S. Agarwal}
\affiliation{Department of Physics, Oklahoma State University, Stillwater, Oklahoma 74078, USA}
\author{M. S. Kim}
\affiliation{QOLS, Blackett Laboratory, Imperial College London, SW7 2BW, United Kingdom}

\date{\today}

\begin{abstract}
We show how to prepare and directly measure the squeezed states of nanomechanical oscillators. An intense pulse interacts with a dielectric mirror in a cavity. The quadratic coupling between the optical pulse and the oscillator results in the reduction of a quadrature variance of the massive oscillator. Differently from others, the proposed scheme here requires neither any post-action on the optical field nor a preparation of a nonclassical optical field. Depending on the initial temperature of the system, the squeezing of the variance under the vacuum limit can be achieved only by a few repetitive interactions.
\end{abstract}


\maketitle

Even though a system of a few tens of nanometer scale is small there can still be more than trillions of atoms hence it is a massive system in quantum mechanics. There have been studies on how to achieve a quantum state of an optomechanical system~\cite{kippenberg-review} where optical fields are coupled to nanomechanical mirrors which are treated as oscillators in order to control the mechanical state.  In this aspect, main experimental effort has been concentrated on achieving the ground state of the oscillator and a huge progress has been made~\cite{marquardt-review}. There have also been some theoretical suggestions to achieve nonclassical states in the optomechanical system~\cite{mancini}. Huang and Agarwal\cite{huang-agarwal} have also suggested that the cooling problem can be bypassed by using squeezed light which due to nonlinear radiation pressure coupling gets transferred to the nano mirror.

\begin{figure}[t]
\centerline{\psfig{figure=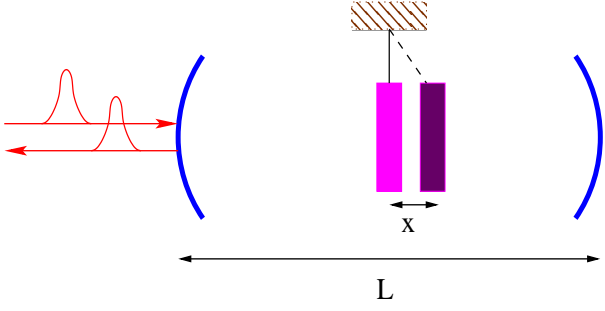,width=8cm,height=4cm}}
\caption{(Color online) Cavity to realize a quaratic kick of the dielectric mirror of reflectivity $R<1$. The nanomechanical oscillator is placed at the antinode of the field, with an antinode in the middle of the cavity.}
\label{fig:setup}
\end{figure}

The nanomechanical oscillator is coupled to an optical field by radiation pressure. Assuming a cavity composed of a high reflectance mirror at one end and a perfect mirror, which is movable, at the other end, the field input to the cavity pushes the movable mirror and changes the cavity length. This causes the change of the cavity mode. When the displacement of the mirror is small, the total process is summarized in the interaction Hamiltonian $H_i=g^\prime\hat{n}\hat{x}$, where $g^\prime$ is the coupling strength and $\hat{n}$ is the number operator for the radiation field and $\hat{x}$ is the dimensionless position operator of the mirror. This is the situation considered for the cooling of a mirror into its ground state. On the other hand, there have recently been works~\cite{harris,harris-b,harris2} on placing a dielectric mirror, whose reflectivity $R<1$, inside a cavity (see Fig.~\ref{fig:setup}). Here, the dielectric mirror is a nanomechanical oscillator interacting with the cavity field. Using this system, nonlinear optomechanical coupling was realized, where the Hamiltonian of the system is~\cite{meystre,agarwal}
\begin{equation}
\hat{H}^\prime=\hat{H}_m+\hat{H}_f+\hbar g\hat{n}\hat{x}^2.
\label{hamiltonian}
\end{equation}
$\hat{H}_m=\hbar {\omega_m\over 2}(\hat{x}^2+\hat{p}^2)$ and $\hat{H}_f=\hbar \omega \hat{a}^\dag\hat{a}$ are the free Hamiltonians for the mechanical oscillator and the field, respectively and the last term is the interaction Hamiltonian showing the quadratic coupling of the optical field with the oscillator's position. The coupling coefficient $g=\frac{2\hbar\omega^2}{m\omega_m Lc}\sqrt{\frac{R}{1-R}}$ where $L$ is the cavity length, $m$ is the mass of the oscillator and $\omega$ and $\omega_m$  the angular frequencies of the optical field and the mechanical oscillator respectively.

In this paper, inspired by the recent experimental progress~\cite{harris2} in strong purely quadratic optomechanical coupling, we propose a scheme to generate a squeezed state and to directly measure the squeezing. There have been a few proposals on generating nonclassical states in the nanomechanical systems~\cite{mancini,huang,ruskov,zhang}, which use feedback mechanisms or nonclassical fields of light. The post-action or the nonclassical field generation may be very inefficient. Recently, Vanner {\it et al.} \cite{vanner} developed a scheme to generate and detect squeezed states, using the stroboscopic interaction of a short pulse with the nanomechanical oscillator. In this scheme, the mechanical state is collapsed into a displaced squeezed state by a homodyne measurement of the pulse field after the interaction. The displacement of the squeezed state depends on the homodyne measurement outcome. Differently from these schemes, the scheme presented here requires neither a feedback nor a measurement of the light field. Utilizing the quadratic coupling, we achieve the squeezing without any action after the interaction. We also show that the quadratic coupling enables the direct measure of squeezing as the observable depends on $\hat x^2$. These are important advantages to show the nonclassical nature of the nanomechanical oscillators. In particular, while probing nonclassicality has been a highly nontrivial task in such systems, we provide a realistic scheme to probe the quadrature variance of the state with minimum resources.

{\it Squeezed state generation}.- In the frame rotating at the frequency of the optical field, the Hamiltonian (\ref{hamiltonian}) becomes
\begin{equation}
\hat{H}=\hbar {\omega_m\over 2}(\hat{x}^2+\hat{p}^2)+\hbar g\hat{n}\hat{x}^2,
\label{hamiltonian2}
\end{equation}
where the explicit form of the oscillator free Hamiltonian is written. In order to find the evolution of the quadrature operators for the mechanical oscillator, we rewrite (\ref{hamiltonian2}) as
\begin{equation}
\hat{H}=\hbar {\omega_m\over 2}\hat{p}^2+\hbar {\tilde{g}\over 2}\hat{x}^2,\ \tilde{g}=2g\hat{n}+\omega_{m}.
\label{hamiltonian3}
\end{equation}
Note that $\hat{n}$ is a constant of motion. The Heisenberg equations of motion $\dot{\hat{x}}=\hat{p}\omega_{m}$, $\dot{\hat{p}}=-\tilde{g}\hat{x}$ can be easily integrated leading to
\begin{equation}
\mbox{e}^{i\hat{H}t/\hbar}\left(\begin{array}{c} \hat{p}\cr \hat{x}\end{array}\right)\mbox{e}^{-i\hat{H}t/\hbar} =K(t) \left(\begin{array}{c}\hat{p} \cr \hat{x}\end{array}\right),
\label{evolution}
\end{equation}
where the transformation matrix due to the kick \cite{Several} by the optical field is
\begin{equation}
K(t)=\left(\begin{array}{cc} \cos\sqrt{\tilde{g}\omega_m}t & -\sqrt{\tilde{g}\over\omega_m}\sin\sqrt{\tilde{g}\omega_m}t\cr
\sqrt{\omega_m\over\tilde{g}}\sin\sqrt{\tilde{g}\omega_m}t  &    \cos\sqrt{\tilde{g}\omega_m}t \end{array} \right).
\label{kick-matrix}
\end{equation}
The transformation matrix depends on the number operator for the cavity field. It is interesting to note that the transformation matrix shows elliptical behavior. At $\sqrt{\tilde{g}\omega_{m}}t=\pi/2$, while $\hat{p}$ becomes $\hat{x}$, with a scaling factor $\sqrt{\tilde{g}/\omega_{m}}$ as well as $\hat{x}$ becomes $\hat{p}$ with a scaling factor $\sqrt{\omega_{m}/\tilde{g}}$. As the product of the scaling factors is 1, the product of the quadrature uncertainties does not change, but these scaling factors would result in squeezing of the mechanical oscillator.

It is natural to assume that the oscillator is at thermal equilibrium before it interacts with an optical field. Thus its mean displacement is zero and $\langle\{\hat{x},\hat{p}\}\rangle=0$, where $\{~,~\}$ is the anticommutator.  The mean variances are $\langle\Delta\hat{p}^2\rangle=\langle\Delta\hat{x}^2\rangle=\bar{n}+1/2$ where the mean excitation number $\bar{n}=1/(\mbox{e}^{\hbar\omega_m/k_BT}-1)$ for a given temperature $T$. $k_B$ is the Boltzmann constant.  The state is called squeezed when the variance is below the zero temperature limit. Now, let us consider an optical pulse with an average photon number around $10^{11}$ and a narrow bandwidth, $\Delta \omega\ll\omega$  in the cavity whose decay rate $\kappa\approx 10^7$sec$^{-1}$. The pulse can thus be considered quasi-monochromatic and the optical operator $\hat{n}$ can be replaced by the average photon number $\bar n_p$. Note that $10^{11}$ cavity photons imply $\kappa\bar{n}=10^{18}$ photons per sec which means the power of about 100mW for the frequency regime we are interested in. We then take the parameters \cite{harris-theory} in the range: $g=10^{-4}$sec$^{-1}$ and $\omega_m=10^6$sec$^{-1}$. The mass of the nanomechanical oscillator is assumed 1ng. The wavelength of the optical field is $\lambda=2\pi c/\omega=532$nm. The cavity length is $L=6.7$cm. The reflectivity of the dielectric mirror is $R\approx0.4$.

From Eq.~(\ref{evolution}) it is clear that the momentum variance is minimum at the initial thermal equilibrium state as $\tilde{g}/\omega_m$ is always greater than $1$ for nonzero coupling $g$. On the other hand, the variance of position is reduced from the initial value  proportional to $\omega_m/\tilde{g}$.  If the pulse interaction time is $\sqrt{\tilde{g}\omega_m}t=\pi/2$, the position variance after the interaction is $\omega_m/\tilde{g}$.  For $g= 10^{-4}$sec$^{-1}$and the parameter values given above, the position variance is reduced by $20$ times, which means that the mechanical state is squeezed below the quantum limit by one interaction when the oscillator is initially in equilibrium with a thermal bath of temperature $T=0.1$mK ($\bar{n}=13$ for $\omega_m=10^6$sec$^{-1}$). In fact, by repetitive interactions between optical pulses and the oscillator, we can further reduce the variance. For instance, when the temperature is $1$mK, the thermal excitation number is $138$ which can be brought down to sub-vacuum level by two pulse interactions.

After an interaction, the optical pulse leaves the cavity and the oscillator is subject to its free Hamiltonian. Using the transformation matrix showing the free evolution of the oscillator,
\begin{equation}
M_f(\tau)=\left(\begin{array}{cc} \cos\omega_m \tau & -\sin\omega_m \tau \cr \sin\omega_m \tau & \cos\omega_m \tau \end{array}\right),
\label{free-matrix}
\end{equation}
the repetitive interaction of the oscillator with $n$ pulses is described by
\begin{equation}
\left(\begin{array}{c} \hat{p} \cr \hat{x}\end{array}\right)_{t,\tau}=K_n(t)M_f(\tau)\cdots K_2(t) M_f(\tau)K_1(t)\left(\begin{array}{c}\hat{p}\cr\hat{x}\end{array}\right),
\label{rep-kick}
\end{equation}
which shows free evolution of the mechanical oscillator between pulse interactions.
Here, we have assumed that the pulse interactions are the same and the pulse intervals are the same. If the free evolution time between the pulse interactions is a quarter of the mechanical oscillator period, the quadrature variance will be reduced efficiently because the free evolution matrix (\ref{free-matrix}) becomes antidiagonal. The impact of error in the free evolution time is as follows. Assume that the pulse interaction time is chosen such that only the off-diagonal terms in Eq.(\ref{kick-matrix}) survive as discussed above, i.e. $\sqrt{\tilde{g}\omega_{m}}t=\pi/2$. After the second pulse interaction with a period of free evolution for a period $\tau$ between the two pulses, then using the transformation matrix (\ref{rep-kick}), we find the quadrature variance as follows:
\begin{equation}
\langle\Delta\hat{p}^2(\tau)\rangle=\langle(\cos\omega_{m}\tau\hat{p}(0)-\frac{\tilde{g}}{\omega_{m}}\sin\omega_{m}\tau\hat{x}(0))^{2}\rangle,
\label{variancep}
\end{equation}
where $\hat{x}(0)$ and $\hat{p}(0)$ are the operators for the initial state of the oscillator. If the oscillator is initially in the thermal state, the variance becomes $(\cos^{2}\omega_{m}\tau+\frac{\tilde{g}^{2}}{\omega_{m}^{2}}\sin^{2}\omega_{m}\tau)(\bar{n}+1/2)$. As mentioned earlier, $\omega_{m}\tau=\pi/2$ gives the optimum reduction of the variance though any error in $\tau$ would affect the value of the variance.

During the free evolution, the mechanical oscillator dissipates in the thermal environment of the given temperature. If the dissipation process is Markovian, the quadrature variance after the thermal interaction of time $\tau$ appears as the sum of the initial variance $\langle\Delta \hat{x}^2\rangle$ and the variance $(\bar{n}+1/2)$ of the thermal environment \cite{kim}:
\begin{equation}
\langle\Delta\hat{x}^2(\tau)\rangle=\mbox{e}^{-\gamma \tau}\langle\Delta \hat{x}^2(0)\rangle+(1-\mbox{e}^{-\gamma \tau})(\bar{n}+1/2),
\label{sum}
\end{equation}
where $\gamma$ is the energy dissipation rate of the mechanical oscillator. If we wait for the duration of $\tau=\pi/\omega_m$, considering a usual value of the dissipation rate $\gamma=0.1$Hz \cite{harris-theory}, the value of the second term is \{$4\times 10^{-2}, 4\times 10^{-5}, 4\times 10^{-6}$\} for the temperatures \{1K, 1mK, 0.1mK\}.  It is clear that at the cryogenic temperature of mK range, the thermal influence during the free evolution is negligible.

{\it Measurement of squeezing}.- There have been only a few practical schemes to read the optomechanical state. Vanner {\it et al.} considered the homodyne measurement of the optical field after a pulse interaction with a nanomechanical oscillator for the full tomography of the oscillator state~\cite{vanner}. There have also been a proposal for a direct probe of the mechanical state using its interaction with a qubit~\cite{t}. We show a scheme to directly probe the degree of quadrature squeezing for the mechanical oscillator.

Consider the interaction of the system with a weak probe pulse after switching off the pump pulse which was used to create squeezing. The interaction Hamiltonian is now
\begin{equation}
\hat{H}=\hbar\omega_c \hat{c}^\dag\hat{c}+\hbar g \hat{c}^\dag\hat{c}\hat{x}^2+\hat{H}_{m}+i\hbar ({\cal E}_p(t)\hat{c}^\dag\mbox{e}^{-i\omega_p t}-\mbox{c.c.}).
\label{measure-hamiltonian}
\end{equation}
Here ${\cal E}_p$ is the probe electric field which drives the cavity field denoted by the cavity operators $\hat{c}$ and $\hat{c}^{\dag}$. c.c stands for the complex conjugate. The field ${\cal E}_p$ has the units of frequency. Let us work again in a frame rotating with the frequency of the probe pulse. The Langevin equation for the cavity field has the form
\begin{equation}
{d\hat{c}\over dt}=-(\kappa +i(\omega_c-\omega_p)+g\hat{x}^2)\hat{c}+{\cal E}_p+\sqrt{2\kappa}\hat{c}_{in},
\label{dynamic}
\end{equation}
where $\hat{c}_{in}$ represents the vacuum field. To lowest order in $g$, we can write the solution as
\begin{equation}
\hat{c}^{(0)}(t)=\int_{t_o}^t {\cal E}_p(\tau)\mbox{e}^{-(\kappa+i(\omega_c-\omega_p))(t-\tau)}d\tau+\cdots,
\label{solution}
\end{equation}
where $t_0$ is the time the probe is switched on. The terms omitted at the end represent the contribution from $\hat{c}_{in}$. However such terms would not contribute to $\langle \hat{c}^{(0)^\dag}(t)\hat{c}^{(0)}(t)\rangle$. The solution of Eq. (\ref{dynamic})  for the  output field first-order in $g$ directly gives the momentum information:
\begin{equation}
\hat{c}^{(1)}(t)=g\int_{t_1}^t\mbox{e}^{-(\kappa+i(\omega_c-\omega_p))\tau}\hat{x}^2(t-\tau)\hat{c}^{(0)}(t-\tau)d\tau.
\label{first-order}
\end{equation}
Note that the free evolution of $\hat{x}^2$ would give terms like $\mbox{e}^{\pm 2i\omega_m t}$ and a d.c. term. If we assume that $\kappa$ is large and in fact bigger than 2$\omega_m$, we can approximate Eq.(\ref{first-order}) by
\begin{equation}
\hat{c}^{(1)}(t)\approx {g \over\kappa}\hat{x}^2(t)\hat{c}^{(0)}(t).
\label{first-order-approx}
\end{equation}
The output field from the cavity is
\begin{equation}
\hat{c}_{out}(t)\approx \hat{c}^{(0)}(t)+{g\over\kappa}\hat{x}^2(t)\hat{c}^{(0)}(t),
\label{out}
\end{equation}
and hence
\begin{eqnarray}
I_{out}(t) &=& \langle\hat{c}_{out}^\dag(t)\hat{c}_{out}(t)\rangle\nonumber \\
&=& \langle \hat{c}^{(0)^\dag}(t)\hat{c}^{(0)}(t)\rangle\left[1+{2g\over\kappa}\langle\hat{x}^2(t)\rangle+{\cal O}(g^2) \right].\nonumber\\
\label{I-out}
\end{eqnarray}
This is our key result on the detection of the quantized motion of the mirror. The output field directly gives motional information. The intensity of the output field is directly proportional to the variance of the position of the mechanical oscillator. It is remarkable that we do not need to calculate the value of the variance on the measured data. The mean value of the data directly reflects the variance in our case. This has been possible because of the second-order coupling in the Hamiltonian (\ref{measure-hamiltonian}).

{\it Remarks}.- We have proposed a scheme to generate a nonclassical state in a massive optomechanical system using the quadratic coupling between optical pulses and a nanomechanical oscillator. It is remarkable that the oscillator prepared in a cryogenic temperature can be squeezed under the quantum limit only by a couple of pulse interactions. It is due to the fact that the degree of squeezing offered in our scheme is so large. The quadratic coupling is also shown to directly measure the variance reduction. We have shown that the proposed experiment is feasible by the current state-of-the-art.

{\it Acknowledgements.}- This work was supported by the UK EPSRC and the NSF Grant No. PHYS 0653494.


\begin{thebibliography}{99}
\bibitem{Several} Several papers discuss the possibility of the production of squeezing using modulation of harmonic oscillator \cite{Agarwal1,Amit}. The first experimental demonstration was using phonons in solids \cite{Merlin}.
\end{thebibliography}
\end{document}